# Higher Accuracy for Bayesian and Frequentist Inference: Large Sample Theory for Small Sample Likelihood

M. Bédard, D. A. S. Fraser and A. Wong

*Abstract.* Recent likelihood theory produces $p$-values that have remarkable accuracy and wide applicability. The calculations use familiar tools such as maximum likelihood values (MLEs), observed information and parameter rescaling. The usual evaluation of such $p$-values is by simulations, and such simulations do verify that the global distribution of the $p$-values is uniform$(0, 1)$, to high accuracy in repeated sampling. The derivation of the $p$-values, however, asserts a stronger statement, that they have a uniform$(0, 1)$ distribution conditionally, given identified precision information provided by the data. We take a simple regression example that involves exact precision information and use large sample techniques to extract highly accurate information as to the statistical position of the data point with respect to the parameter: specifically, we examine various $p$-values and Bayesian posterior survivor $s$-values for validity. With observed data we numerically evaluate the various $p$-values and $s$-values, and we also record the related general formulas. We then assess the numerical values for accuracy using Markov chain Monte Carlo (McMC) methods. We also propose some third-order likelihood-based procedures for obtaining means and variances of Bayesian posterior distributions, again followed by McMC assessment. Finally we propose some adaptive McMC methods to improve the simulation acceptance rates. All these methods are based on asymptotic analysis that derives from the effect of additional data. And the methods use simple calculations based on familiar maximizing values and related informations.

The example illustrates the general formulas and the ease of calculations, while the McMC assessments demonstrate the numerical validity of the $p$-values as percentage position of a data point. The example, however, is very simple and transparent, and thus gives little indication that in a wide generality of models the formulas do accurately separate information for almost any parameter of interest, and then do give accurate $p$-value determinations from that information. As illustration an enigmatic problem in the literature is discussed and simulations are recorded; various examples in the literature are cited.

Mylène Bédard is Assistant Professor, Département de Mathématiques et de Statistique, Université de Montréal, C.P. 6128, succ. Centre-ville, Montréal, Québec, Canada H3C 3J7 e-mail: bedard@dms.umontreal.ca. D. A. S. Fraser is Professor, Department of Statistics, University of Toronto, 100 St George St., 6th floor, Toronto, Ontario, Canada M5S 3G3 e-mail: dfraser@utstat.toronto.edu. A. Wong is Professor, Department of Mathematics and Statistics,

York University, 4700 Keele Street, Toronto, Ontario, Canada M3J 1P3 e-mail: august@yorku.ca.

## 1. INTRODUCTION

We explore various large sample and likelihood methods for obtaining Bayesian and frequentist $p$-values from a regular statistical model and data. Numerical values are obtained for a simple example that indicates the ease with which the methods can be applied, given the typically available maximum likelihood and related calculations. The general formulas are presented and discussed.

The example, ignoring the nonnormality of error, is simple and transparent: one could plot the data, calculate means and standard deviations, do the bootstrap, or even record likelihood and get broadly about the same answer. But the large sample techniques, more exactly data-accretion techniques, provide accurate separation of component parameter information, precisely summarize the available information, and give accurate determinations of corresponding $p$-values.

In Section 2 we take a pragmatic approach and obtain $p$-values using simple departure measures and distributional approximations related to the Central Limit Theorem. Then in Section 3 we formally reference the statistical model and obtain a $p$-value based on the signed likelihood root. In Section 4 we add a widely accepted default prior and obtain the posterior survivor value, the analogue of the frequentist $p$-value. For the example these require three-dimensional integration.

But then in Section 5 we examine recently developed likelihood-based approximations that have third-order accuracy; numerical values are obtained for the example, and the general formulas are discussed. In Section 6 we discuss the corresponding frequentist third-order $p$-values. Numerical values for the example are then presented together with various intermediate values that indicate how the calculations proceed.

In Sections 7 and 8 we consider exact $p$-values for the preceding methods, as derived by Markov chain Monte Carlo. As part of this we note that a $N = 4 \times 10^6$ simulation in the particular context gives about two-figure accuracy for probabilities, about the same as the third-order approximation methods.

In Section 9 we briefly discuss the role of precision information in the Bayesian and frequentist contexts. Section 10 looks directly at Bayesian means and variances and how they can be approximated by recent likelihood-based methods. Again Markov chain Monte Carlo is used to evaluate the accuracy.

Section 11 presents some intuitive thoughts on the Metropolis–Hastings step in Markov chain Monte Carlo and then proposes several asymptotic and adaptive modifications of the direct McMC approach; these are explored for $p$-values in Sections 12 and 13. A controversial example is examined in Section 14, and some concluding remarks are recorded in a discussion Section 15.

The Bayesian and frequentist methods give about the same answer for the example. In fact, for the particular example they give theoretically the same answer, a consequence of the judicious choice of default prior for the Bayesian analysis. We do not address here the manner of making such judicious choices or how the choice typically needs to be targeted on the particular parameter of interest; this will be addressed subsequently.

## 2. A SIMPLE EXAMPLE: DEPARTURE OF DATA FROM PARAMETER VALUE

Consider an example to illustrate the formulas coming from large sample or, more exactly, data-accretion techniques: a small data set involving a response $y$ with possible linear dependence on a related variable $x$:

(1)

| $x$ | $-3$ | $-2$ | $-1$ | $0$ | $1$ | $2$ | $3$ |
|---|---|---|---|---|---|---|---|
| $y$ | $-2.68$ | $-4.02$ | $-2.91$ | $0.22$ | $0.38$ | $-0.28$ | $0.03$ |

The response variability is taken to be thicker tailed than the normal, say the frequently suggested Student(7) distribution. Then with linear dependence and constant response variability, we have the model

$$f(y;\theta) = \sigma^{-7} \prod_{i=1}^{7} h\{\sigma^{-1}(y_i - \alpha - \beta x_i)\},$$

where $h(z) = \{\Gamma(4)/\Gamma(1/2)\Gamma(7/2)\sqrt{7}\}(1 + z^2/7)^{-4}$ is the Student(7) density. Or using quantile function form, we can write $y_i = \alpha + \beta x_i + \sigma z_i$, where the latent errors $z_i$ are independent Student(7). Now suppose we are interested in assessing the response dependence on $x$ as given by the regression parameter $\beta$, with particular interest in whether $\beta = 1$.

As background we note that the response data were in fact generated from the given model with Student(7) error and then rounded to two decimal places; the parameter values used to generate the data were $\alpha = 0$, $\beta = 1$ and $\sigma = 1$. In passing we note that $\sigma$ is an error scaling and does not record directly the error standard deviation. Also there is no implied connection between the number of observations $n = 7$ and the degrees of freedom for the

HIGHER ACCURACY FOR BAYESIAN AND FREQUENTIST INFERENCE 3error density $df = 7$. The example has simple and transparent structure, and we use it to examine various frequentist and Bayesian assessment methods; we then apply Markov chain Monte Carlo sampling to check the distributional validity of the resulting values.

The example is simple and transparent and we could probably do as well by plotting or by least squares and standard deviations. The likelihood model theory, however, gives a precise separation of information concerning parameters of interest and an accurate determination of the $p$-value or percentage position of data with respect to a value for the parameter of interest. As a more general illustration, we later record simulation data for a challenging example from the literature. We also cite various examples that have been examined in the literature.

For the example considered, a pragmatic first step is to use least squares to separate out the general location and the linear dependence on the related variable $x$. The fitted values for $\alpha$ and $\beta$ are

$$a = -1.322857, \quad b = 0.675000,$$

with residual length $s = (\text{SSR})^{1/2} = 2.660046$ obtained from the sum of squares of residuals.

The parameter $\beta$ records the indicated linear dependence of $y$ on $x$. To assess the value $\beta = 1$ in the presence of the data, we can reasonably examine the raw departure $b - \beta = -0.325000$, and then standardize it by its estimated standard deviation to obtain a standardized measure of departure of data from parameter value, giving

$$(2) \quad \begin{aligned} t &= \frac{b(y) - \beta}{s/\sqrt{5}\sqrt{28}}, \\ t^0 &= \frac{0.675000 - 1}{2.660046/\sqrt{5}\sqrt{28}} = -1.445634. \end{aligned}$$

To interpret this statistically we need information concerning the distribution of possible values for $t$ in the context where the true value of $\beta$ is 1. To this end, some reference to large sample theory suggests the use of the standard normal distribution function, say $\Phi(\cdot)$. The observed value of this distribution function then gives an approximation to the percentage of possible values of $t$ that would be less than the observed $t^0$, in other words, to the percentage position of the data with respect to the hypothesized value $\beta = 1$; this is called the observed $p$-value. Using the standard normal then as an approximation, we obtain the approximate $p$-value

$$(3) \quad \begin{aligned} p_N &= \Phi(t^0) = \Phi(-1.445634) \\ &= 0.07414 = 7.414\%, \end{aligned}$$

which records the observed level of significance in an elemental form, as just the percentage position of the data point or the probability left of the data point, under the hypothesis.

A simple modification hopefully to accommodate the estimation of error scaling is obtained by using the Student(5) distribution function, say $H_5(\cdot)$, as a revised approximation method. We then obtain the approximate $p$-value

$$(4) \quad \begin{aligned} p_S &= H_5(t^0) = H_5(-1.445634) \\ &= 0.10395 = 10.395\%. \end{aligned}$$

An alternative to the direct use of the large sample distribution theory is provided by the bootstrap approach. Using the least squares values, we separate the data values into a location or fit component $\hat{y}_i$ and a residual component $y_i - \hat{y}_i$:

| $x$ | $-3$ | $-2$ | $-1$ | $0$ | $1$ | $2$ | $3$ |
|---|---|---|---|---|---|---|---|
| $\hat{y}$ | $-3.3479$ | $-2.6729$ | $-1.9979$ | $-1.3229$ | $-0.6479$ | $0.0271$ | $0.7021$ |
| $y - \hat{y}$ | $0.6679$ | $-1.3471$ | $-0.9121$ | $1.5429$ | $1.0279$ | $-0.3071$ | $-0.6721$ |

For the bootstrap we randomly sample the residuals with equal probability and add them back to the location component, thus obtaining a bootstrap data set, from which we calculate the bootstrap $t$-statistic value

$$t^* = \frac{b^* - 1}{s^*/\sqrt{5}\sqrt{28}},$$

where $b^*$ and $s^*$ are the regression coefficient and residual length from the bootstrap sample. We repeated this bootstrap sampling a convenient total of $N = 10{,}000$ times, and the empirical distribution function was evaluated at the observed $t^0 = -1.445634$. This gave an observed bootstrap $p$-value:

$$(5) \quad p_{\text{BS}} = \hat{F}(t^0) = 0.1051 = 10.51\%,$$

where $\hat{F}(t)$ is the empirical distribution function of the bootstrap $t^*$ values.

With the present small sample size $n = 7$ we can calculate the bootstrap $p$-value exactly, $p_{\text{ExBS}}$, by using equal probability for each of the $7^7 = 823{,}543$ possible bootstrap samples. We then take

$$(6) \quad \begin{aligned} p_{\text{ExBS}} = &\{\text{proportion}(t^* < t^0) \\ &+ \text{proportion}(t^* \leq t^0)\}/2, \end{aligned}$$

TABLE 1
*Simple frequentist p-values for the regression example in Section 2*

| Measure of departure | Distributional approximation | p-value $\beta=1$ | $\beta=1.5$ | $\beta=2$ |
|---|---|---|---|---|
| $t$-statistic | Normal | 0.07414 | $0.0^3 121$ | $0.0^8 189$ |
| $t$-statistic | Student(5) | 0.10395 | $0.0^2 722$ | $0.0^2 100$ |
| $t$-statistic | Bootstrap $N=10^4$ | 0.10750 | $0.0^2 790$ | $0.0^3 800$ |
| $t$-statistic | Bootstrap (exact) $N=7^7$ | 0.10332 | $0.0^2 833$ | $0.0^3 888$ |
| SLR | Normal | 0.05774 | $0.0^2 148$ | $0.0^4 830$ |

A $p$-value records the percentage position of the data relative to a possible true value $\beta$ for the parameter; $\beta=1$ is in fact the true value underlying the data set; $\beta=1.5$ and $\beta=2$ are other values that might have been of interest. Multiple zeros are indicated by a superscript, thus $0.0^2 148 = 0.00148$.

called a mid-$p$-value, and obtain

$$p_{\text{ExBS}} = 0.1033231 = 10.332\%; \tag{7}$$

with our particular rounded $t^0$ value there were no values at the boundary point. These four approximate $p$-values make use of a pragmatic choice of departure measure combined with distributional information provided by Central Limit Theorem-type approximations or by resampling from the nonparametric maximum likelihood distribution; and they provide us with four determinations of the observed percentage position of the data relative to the model with $\beta=1$. Other initial departure measures could have been considered, as well as other distributional approximations or determinations.

These $p$-values for testing $\beta=1$ are recorded in Table 1 together with a likelihood-based method discussed in the next section. The table also records $p$-values for testing the more extreme $\beta$ values, 1.5 and 2, with corresponding observed values $t^0 = -3.669685$ and $t^0 = -5.893737$.

## 3. THE EXAMPLE: SIMPLE LIKELIHOOD DEPARTURE MEASURE

More intrinsic departure measures are available from long available likelihood theory. The likelihood function in the present Student regression context is

$$L(\alpha, \beta, \sigma; y) = c\sigma^{-7} \prod_{i=1}^{7} \left\{ 1 + \frac{(y_i - \alpha - \beta x_i)^2}{7\sigma^2} \right\}^{-4}, \tag{8}$$

with log-likelihood

$$\ell(\alpha, \beta, \sigma; y)$$

$$= a - 7\log\sigma \tag{9}$$

$$- 4 \sum_{i=1}^{7} \log\left\{ 1 + \frac{(y_i - \alpha - \beta x_i)^2}{7\sigma^2} \right\};$$

the observed likelihood $L^0(\alpha, \beta, \sigma)$ and observed log-likelihood $\ell^0(\alpha, \beta, \sigma)$ are obtained by substituting the data $y^0 = (y_1^0, \ldots, y_7^0)$ from the data array (1) into the above expressions.

In a general context the observed likelihood function is given as

$$L^0(\theta) = L(\theta; y^0) = cf(y^0; \theta) = f^0(\theta),$$

which is the observed density function, that is, the statistical model $f(y; \theta)$ examined at the observed data point $y^0$; it records the amount of probability sitting at that data point, viewed as a function of possible values for the parameter. The constant $c$ is taken as arbitrary but positive and indicates that only relative values from one $\theta$ value to another are of relevance given the data point. If we consider in general how likelihood depends on data we can write

$$L(\theta; y) = cf(y; \theta),$$
$$\ell(\theta; y) = a + \log f(y; \theta),$$

where $c > 0$ and $a, c$ are otherwise arbitrary for each choice of data point $y$.

Suppose now that we are interested in a scalar component $\psi = \psi(\theta)$. Most likelihood methods make use of maximum likelihood values (MLEs); we do, however, avoid referring to them as estimates, as they are useful but typically not directly as estimates. We write $\hat{\theta} = \arg\sup L(\theta)$ for the value that maximizes $L(\theta)$. Also if $\psi(\theta)$ is a component parameter of particular interest, we write $\hat{\theta}_\psi = \arg\sup_{\psi(\theta)=\psi} L(\theta)$ for the value that maximizes $L(\theta)$ subject to the interest parameter $\psi(\theta)$ having some value $\psi(\theta) = \psi$ of special interest. Then based on the likelihood $L(\theta)$ alone, an important departure measure of data from $\psi(\theta) = \psi$ is obtained as the signed likelihood root (SLR)

$$r_\psi = \text{sign}(\hat{\psi} - \psi)[2\{\ell(\hat{\theta}) - \ell(\hat{\theta}_\psi)\}]^{1/2}; \tag{10}$$

this measure examines how probability at the data point under the full model exceeds that when $\psi(\theta)$ is restricted to the value $\psi$, and theory has shown it to be fundamental. One way of viewing this measure is to picture how much the log-likelihood rises from the maximum when $\psi(\theta) = \psi$ up to the overall maximum when $\theta$ is unrestricted, that is, $\ell(\hat{\theta}) -$ $\ell(\hat{\theta}_\psi)$, often written $\hat{\ell} - \tilde{\ell}$. We take this rise as having quadratic form $r_\psi^2/2$ in terms of some quantity $r_\psi$; solve for $r_\psi$; and then attach the appropriate sign. For testing a true value $\psi(\theta) = \psi$ the signed likelihood root $r_\psi$ has to first order a standard normal limiting distribution, a follow-through from the Central Limit Theorem. Related measures could be based on the slope of the log-likelihood at the tested value or on the displacement from $\hat{\theta}_\psi$ to $\hat{\theta}$, but neither has the same mathematical invariance or the same track record in applications.

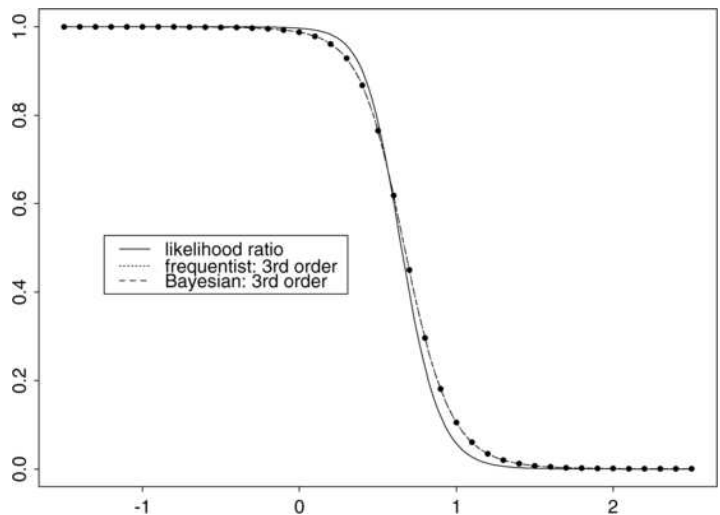

FIG. 1. *For the example in Section 2: the third-order Bayesian survivor function, the third-order frequentist p-value and the first-order SLR p-value.*

For many likelihood calculations, particularly recent higher-order calculations, the computationally challenging aspects often arise in the maximization steps rather than in other steps.

For our simple regression example and the related likelihood calculations we now use maximum likelihood rather than least squares variables; the observed values obtained by computer iteration are

$$\hat{\alpha} = -1.3504512,$$
$$\hat{\beta} = 0.6504019,$$
$$\hat{\sigma} = 0.9641110$$

for the overall MLEs, and

$$\tilde{\alpha} = \hat{\alpha}_{\beta=1} = -1.366699,$$
$$\tilde{\beta} = \hat{\beta}_{\beta=1} = 1,$$
$$\tilde{\sigma} = \hat{\sigma}_{\beta=1} = 1.154527$$

for the constrained MLEs when $\beta = 1$. In passing we note that the use of MLE variables can be convenient but can be awkward when the error distribution itself has dependence on the parameter; for if the distribution for the error itself has dependence on the parameter rather than as here being just Student(7), then the maximum likelihood value could also have that parameter dependence and thus not be a statistic. From the preceding numerical values we obtain from (10) the signed likelihood root

(11) $$r_{\beta=1} = -1.574053.$$

The corresponding observed *p*-value based on the first-order normal approximation for $r$ is then

(12) $$p_{\text{SLR}} = \Phi(r_{\beta=1}) = 5.774\%;$$

this is also recorded in Table 1. In Figure 1 we plot the SLR *p*-value against a full range of possible $\beta$ values; this is called the *p*-value function; some related determinations discussed in later sections are also recorded in the figure. Other likelihood departure measures based on the score and maximum likelihood estimates are sometimes considered, but they frequently have serious distributional and measurement bias difficulties. By contrast, the SLR-based approximate *p*-value uses a departure measure that directly relates to the statistical model; it summarizes background information contained in the model combined with distributional information derived from the large sample behavior of the likelihood function.

## 4. ANALYSIS WITH DEFAULT PRIOR

An alternative first-order likelihood-based method comes from the use of a model-based flat prior called a default prior together with conditional-probability type calculations, typically referred to as Bayes algorithm. Let $\pi(\theta)$ be a weight function for $\theta$ based on symmetries, invariance, or other relevant properties of the model. The corresponding posterior distribution viewed as providing inference information concerning $\theta$ is

$$\pi(\theta|y^0) = c\pi(\theta)L(\theta), \quad (13)$$

where $c$ now indicates the norming constant. This default approach was implicit in Bayes (1763), was strongly promoted by Laplace (1812) and Jeffreys (1946) and acquired the name inverse probability. More recently its development has been stimulated by the Valencia conferences (see, e.g., Bayesian Statistics 7, 2003); the default priors are called objective priors but in such contexts the objective can only refer to model structure, not to any objective frequencies in the context being examined.

For inference concerning an interest parameter, say $\psi(\theta)$, one might then reasonably calculate the marginal posterior density function

$$\pi(\psi|y^0) = \int \pi(\theta|y^0) \, d\lambda,$$

where $\lambda$ is a complementing nuisance parameter here chosen so that $\theta$ is one–one equivalent to $(\psi, \lambda)$ and to have, say, Jacobian $|\partial\theta/\partial\psi\,\partial\lambda| \equiv 1$ so that support volume corrections can be ignored.

This marginalization to obtain an inference distribution for a component parameter can produce misleading results (e.g., Dawid, Stone and Zidek, 1973). To overcome this issue, a preferred approach is to have a prior that depends on the particular parameter of interest, and thus to use a targeted prior $\pi_\psi(\theta)$ where the subscript indicates the particular parameter being targeted (e.g., Jeffreys, 1946; Bernardo, 1979; Fraser et al., 2003). We do not address this important issue of choosing default priors but do acknowledge that it is of major interest for the Bayesian community at the present time and in part for the frequentist community.

For our simple regression example we might possibly consider the model-based default prior $\pi(\theta)$ to be the invariant prior

$$\pi(\theta)\,d\theta = \frac{d\alpha\,d\beta\,d\sigma}{\sigma^3} = \frac{d\alpha\,d\beta\,d\log\sigma}{\sigma^2};$$

this derives from parameter transformations on the sample and parameter spaces and is referred to as the left invariant prior (e.g., Jeffreys, 1946; Fraser, 1979): under transformations that make location and scale changes on the initial sample space, the differential rewritten as, say, $d\alpha\,d\beta\,d\sigma/\sigma^3$ remains unchanged, is invariant. This left prior avoids the marginalization issues for certain parameter components that are linear in a location parameterization implied by asymptotic theory (Fraser and Reid, 2002); for many familiar parameters of interest, however, it can lead to the marginalization issues; and furthermore it does not correspond to the confidence theory pivotal inversion based on the usual equations $(y_i - \alpha - \beta x_i)/\sigma = z_i$.

For the parameter $\beta$, various approaches (e.g., Dawid, Stone and Zidek, 1973; Fraser, 1979) suggest the targeted prior

$$\pi_\beta(\theta)\,d\theta = \frac{d\alpha\,d\beta\,d\sigma}{\sigma} = d\alpha\,d\beta\,d\log\sigma,$$

and for many parameters having a certain linearity it does avoid the marginalization issue; some discussion of this linearity and a related curvature measure may be found in Fraser and Reid (2002); the curvature issue does not arise in the present problem for the nice parameters $\alpha, \beta, \sigma$ and we do not pursue the issue here. From the transformation viewpoint this is called the right invariant prior. The corresponding model-based posterior for $\theta$ is then given by

$$
\begin{aligned}
\pi(\alpha, \beta, \tau; y^0)\,d\theta \\
(14) \quad &= c\sigma^{-7} \\
&\quad \cdot \prod_{i=1}^{7}\left\{1 + \frac{(y_i^0 - \alpha - \beta x_i)^2}{7\sigma^2}\right\}^{-4} d\alpha\,d\beta\,d\log\sigma;
\end{aligned}
$$

for this, if we now take $\theta$ to be $(\alpha, \beta, \tau) = (\alpha, \beta, \log\sigma)$, the implied prior is $\pi(\theta) = 1$ and it conforms to confidence inversion.

In order then to obtain the marginal density for the interest parameter $\beta$, an integration over $\alpha$ and $\tau$ is required. Repeated numerical integration over two dimensions can be quite feasible but often is not easily implemented; we next consider some alternative integration procedures.

For more general use of this Bayesian approach the choice of the default prior becomes a crucial issue and is the focus of much present activity in the Bayesian community; the term objective Bayesian prior is sometimes used in place of the term default

Bayesian prior, but this is misleading as the objective would indicate that it is describing the physical context rather than being based as here on model characteristics, a level removed from the physical context.

## 5. THIRD ORDER WITH DEFAULT PRIOR

For many regular densities, posterior or otherwise, the Laplace integration method provides an accurate alternative route for obtaining the marginal density of a component, such as $\beta$ in our case. As before, but in general notation, we use $\pi(\psi, \lambda)$ for the proposed prior and $L(\psi, \lambda)$ for the likelihood. Then with third-order accuracy, the marginal posterior density for $\psi$ can be obtained by Laplace integration over the nuisance parameter divided by Laplace integration over the full parameter, giving

$$\pi(\psi; y^0) = \frac{e^{k/n}}{(2\pi)^{d/2}} \tag{15}$$
$$\cdot e^{-r_\psi^2/2} \left\{ \frac{|\hat{j}_{\theta\theta}|}{|j_{\lambda\lambda}(\hat{\theta}_\psi)|} \right\}^{1/2} \frac{\pi(\hat{\theta}_\psi)}{\pi(\hat{\theta})},$$

where $k$ indicates a first-order constant; $r_\psi^2$ is the likelihood ratio quantity

$$r_\psi^2 = 2\{\ell(\hat{\theta}) - \ell(\hat{\theta}_\psi)\} \tag{16}$$

discussed earlier but now used more generally with $\psi$ of dimension $d$, nuisance parameter $\lambda$ of dimension $m$, and $p = d + m$; the Hessian matrices

$$j_{\theta\theta}(\theta; y) = -\frac{\partial^2}{\partial\theta\,\partial\theta'}\ell(\theta; y),$$

$$j_{\lambda\lambda}(\theta; y) = -\frac{\partial^2}{\partial\lambda\,\partial\lambda'}\ell(\theta; y)$$

are information functions for the full and for the nuisance parameters, and have dimensions $p \times p$ and $m \times m$; they are just negative second-derivative matrices of the log-likelihood function, and when evaluated at $\hat{\theta}$ and $\hat{\theta}_\psi$ give the observed information matrices

$$\hat{j}_{\theta\theta} = j_{\theta\theta}(\hat{\theta}; y), \tag{17}$$
$$\tilde{j}_{\lambda\lambda} = j_{\lambda\lambda}(\hat{\theta}_\psi) = j_{\lambda\lambda}(\hat{\theta}_\psi; y).$$

Numerically, the information matrices can typically be computed by taking second differences based on very small and equally spaced values for each coordinate.

The preceding marginal density can also be written

$$\pi(\psi; y^0) = \frac{e^{k/n}}{(2\pi)^{d/2}} \frac{L_P(\psi)}{L_P(\hat{\psi})} \tag{18}$$
$$\cdot \left\{ \frac{|\hat{j}_{\theta\theta}|}{|j_{\lambda\lambda}(\hat{\theta}_\psi)|} \right\}^{1/2} \frac{\pi(\hat{\theta}_\psi)}{\pi(\hat{\theta})},$$

where

$$L_P(\psi) = \sup_\lambda L(\psi, \lambda) = L(\psi, \hat{\lambda}_\psi)$$

is the profile likelihood function for $\psi$, obtained by maximizing the full likelihood over $\lambda$ for fixed value $\psi$ of the interest parameter $\psi(\theta)$.

The methods inherent in the Laplace integration procedure can be described fairly easily. A regular function, here $L(\psi, \lambda)$ for fixed $\psi$, whose logarithm has additive and maximum likelihood value properties under increasing sample size $n$ can be rewritten

$$f(\lambda) = e^{h(\lambda)}$$
$$= c \exp\{-\hat{j}_{\lambda\lambda}\lambda^2/2\} \exp\{a\lambda^3/6n^{1/2} + b\lambda^4/24n\}$$

to third order as a function of $\lambda$, with obvious generalization for vector $\lambda$; for this we are letting $\lambda$ designate a standardized departure of the original $\lambda$ from the maximizing value for $f(\lambda)$ with $\psi$ fixed. After expanding the second exponential in a power series and similarly for the log-prior, and then integrating term by term with respect to $\lambda$, we obtain to fourth order,

$$\int f(\lambda)\pi(\psi, \lambda)\,d\lambda$$
$$= e^{k/n}(2\pi)^{m/2}|\hat{j}_{\lambda\lambda}(\psi)|^{-1/2}f(\hat{\lambda}_\psi)\pi(\psi, \hat{\lambda}_\psi),$$

where $\hat{j}$ is the negative Hessian with respect to $\lambda$ as evaluated at the maximum for the fixed $\psi$. The integrations are based on simple reference to the multivariate normal integral; for some background, see Strawderman (2000) and for some discussion of term by term integration, see Andrews, Fraser and Wong (2005).

For a scalar interest parameter $\psi$ we can reasonably be more interested in an integral of its density function, and particularly in the right tail integral called the posterior survivor function. Why the right tail? Consider the simple case of a variable $x$ measuring a parameter $\psi$ with error density $f(e)$ and distribution function $F(e)$; we have: the observed p-value is $p^0(\psi) = F^0(\psi) = F(x^0 - \psi)$; the right tail posterior survivor function with a natural flat prior is $s(\psi) = \int_\psi^\infty f(x^0; \alpha) \, d\alpha = \int_\psi^\infty f(x^0 - \alpha) \, d\alpha = F(x^0 - \psi)$; and these are equal. In more general model situations the $p$-value as discussed in the next section records in a statistical sense where the data value is with respect to $\psi$ in a left-to-right distributional sense and then corresponds as in the simple case to the survivor posterior value $s(\psi)$.

For the general case a highly accurate approximation to the posterior survivor function is available from likelihood theory (see, e.g., Fraser, Reid and Wu, 1999, generalizing DiCiccio and Martin, 1991):

$$
\begin{aligned}
(19) \quad s(\psi) &= 1 - G(\psi|y^0) \\
&= \Phi\left(r - r^{-1} \log \frac{r}{q}\right) = \Phi(r^*).
\end{aligned}
$$

For this, $G$ designates the posterior distribution function for $\psi$, $r$ is the signed likelihood root $r_\psi$ given as (10) in Section 3, $q_B$ is a score-type departure measure for $\psi$,

$$
(20) \quad q_B = \ell_\psi(\hat{\theta}_\psi) \left\{ \frac{|\hat{j}_{\lambda\lambda}(\hat{\theta}_\psi)|}{|\hat{j}_{\theta\theta}|} \right\}^{1/2} \frac{\pi(\hat{\theta})}{\pi(\hat{\theta}_\psi)},
$$

where

$$
(21) \quad \ell_\psi(\theta) = \frac{\partial}{\partial \psi} \ell(\hat{\theta}_\psi) = \frac{\partial}{\partial \psi} \ell(\theta; y^0) \Big|_{\hat{\theta}_\psi}
$$

is a score departure measure, and $r^*$ is implicitly defined. Note that for convenience we have taken the full parameter $\theta$ to be given as $(\psi, \lambda')'$ in terms of the components, and this applies as well to the information matrices at (17).

Also we have chosen to record the upper tail for presenting posterior probability from the posterior distribution; the interest parameter will often have physical meaning in a particular application and investigators will think in terms of a variable measuring the parameter as, say, with a maximum likelihood estimate. In such a framework the usual $p$-value is left tail for the variable and correspondingly is right tail for the parameter as in the location case; accordingly for harmony between the two inference approaches we take the reference Bayesian probability to be the upper tail survivor value. In the next section we will find that formula (20) using $r$ from (10) and $q_B$ from (20) can be derived directly from frequentist formulas in the next section by acting as if $\pi(\psi; y^0)$ in (15) were obtained from a location model $\pi(\psi - x; y^0)$ with a nominal variable $x$ taking the observed value $x = 0$.

For our simple example and testing $\beta = 1$, we have $r = -1.574053$ from (11) and we have

$$
(22) \quad q_B = -0.9483686,
$$

where

$$
\ell_\beta(\hat{\theta}_{\beta=1}) = -5.868699,
$$

and the full and constrained information matrices for $\theta = (\alpha, \beta, \tau)$ are

$$
(23) \quad \begin{aligned}
\hat{j}_{\theta\theta} &= j_{\theta\theta}(\hat{\theta}) \\
&= 911 \begin{pmatrix} 5.7894195 & 1.311060 & -0.3286837 \\ 1.311060 & 27.288552 & -1.3395559 \\ -0.3286837 & -1.339556 & 12.1900132 \end{pmatrix},
\end{aligned}
$$

$$
(24) \quad j_{\lambda\lambda}(\hat{\theta}_\beta) = 911 \begin{pmatrix} 4.0240689 & -0.3319537 \\ -0.3319537 & 8.5925687 \end{pmatrix},
$$

with corresponding determinants

$$
(25) \quad \begin{aligned}
|\hat{j}_{\theta\theta}| &= 1892.702, \\
|j_{\lambda\lambda}(\hat{\theta}_\beta)| &= 34.4669.
\end{aligned}
$$

This gives $r^* = -1.252169$; the Bayesian posterior survivor value from (18) is then

$$
(26) \quad s_B(1) = 0.1052542;
$$

this is recorded in Table 2 together with the Bayesian survivor values for testing $\beta = 1.5$ and $\beta = 2$, as well as some McMC validation results discussed in later sections.

The first-order likelihood method in Section 3 requires the full and the constrained maximum likelihood values $\hat{\theta}$ and $\hat{\theta}_\psi$ with of course corresponding values for the log-likelihood function. In order to take advantage of the approximate integration formulas in this section, we require in addition the second-derivative values at each MLE; such derivatives are of course also needed for familiar score and MLE departure measures and typically can be obtained by differencing.

We can also calculate the Bayesian survivor value $s(\beta)$ for a range of values for $\beta$. For our special example the Bayesian survivor function $s_B(\beta)$ is plotted in Figure 1 together with the likelihood ratio $p$-value $\Phi(r_\beta)$ and a third-order frequentist $p$-value to be discussed in the next section.

## 6. THIRD-ORDER $p$-VALUE

Recent likelihood methods give highly accurate approximations for frequentist inference in much the same manner as for the Bayesian context just described. The methods require full and constrained maximum likelihood values $\hat{\theta}$ and $\hat{\theta}_\psi$, as well as full and constrained information determinants. They also, however, need something more in the way of information from the model and data. The nature of this extra information can best be described in terms of parameterization scaling or reexpression. In particular, we need to express the initial parameter $\theta$ as a canonical-type parameter, say $\varphi = \varphi(\theta)$. In the Bayesian context, such additional information is closely related to the development of an appropriate default prior; thus the use of a weighted likelihood $L(\theta)\pi(\theta)$ can partly be interpreted in terms of seeking a location reparameterization $\beta = \beta(\theta)$ such that $\pi(\theta)\,d\theta = d\beta$. Bayesian parameter reweightings have long been sought in the developmental sequence from invariant (Bayes, 1763; Laplace, 1812) to Jeffreys (1946) to reference priors (Bernardo, 1979).

In the frequentist context the accessible reparameterizations are of an exponential rather than a location type, but they give some access to the related location information. The exponential-type reparameterization can be examined initially in the context of an exponential model. To this effect, consider an exponential model with canonical parameter $\varphi$,

$$(27) \quad f(y;\varphi) = \exp\{\varphi' s(y) - \kappa(\varphi)\}h(y),$$

where $\varphi$ and $s$ have the same dimension, say $p$. For such a model, the saddlepoint approximation (Daniels, 1954) can be remarkably accurate, and has the density form

$$(28) \quad \bar{f}(y;\varphi)\,dy = \frac{e^{k/n}}{(2\pi)^{p/2}}\,e^{-r^2/2}|\hat{j}_{\varphi\varphi}|^{1/2}\,d\hat{\varphi}$$

$$(29) \quad = \frac{e^{k/n}}{(2\pi)^{p/2}}e^{-r^2/2}|\hat{j}_{\varphi\varphi}|^{-1/2}\,ds,$$

where $r^2$ is the likelihood ratio quantity for assessing the full parameter $\varphi$ and $\hat{j}_{\varphi\varphi} = -(\partial^2/\partial\varphi^2)\ell(\varphi;y^0)|_{\hat{\varphi}}$ is the observed information matrix from a data point $y$. The renormalized (28) is third-order accurate.

For the important special case of a scalar parameter $\varphi$, a corresponding distribution function approximation was developed by Lugannani and Rice (1980) and in an alternative form by Barndorff-Nielsen (1991). Both versions use the signed likelihood ratio $r = r(\varphi, s)$ corresponding to (10), plus a maximum likelihood value departure $q = q(\varphi, s)$:

$$q_f = (\hat{\varphi} - \varphi)\hat{j}_{\varphi\varphi}^{1/2}.$$

The approximate distribution function in the Barndorff-Nielsen (1991) form for $\hat{\varphi}$ or $s$ is then

$$(30) \quad \bar{F}(s;\varphi) = \Phi\left(r - r^{-1}\log\frac{r}{q_f}\right) = \Phi(r^*),$$

and has third-order accuracy; this has the same form as (19) but uses a different departure $q$ appropriate to the present frequentist context. The similarity of the Bayesian formula (19) and the above frequentist formula can appear more plausible by defining the following reexpressions of the variable and the parameter:

$$\beta(\varphi) = \int^\varphi \hat{j}_{\varphi\varphi}^{1/2}\,d\hat{\varphi},$$
$$b(s) = \int^s \hat{j}_{\varphi\varphi}^{-1/2}\,ds.$$

TABLE 2
*For the simple regression example in Section 2, Bayesian s-values for assessing the values $\beta = 1$, $\beta = 1.5$ and $\beta = 2$: using the third-order formula (19); using the McMC (Section 8 with Normal proposal); using AMcMC (Section 12 with adaptive choice of Student proposal)*

| Test procedure | s-value | | |
|---|---|---|---|
| | $\beta = 1$ | $\beta = 1.5$ | $\beta = 2$ |
| Bayesian: third order ($N = 4 \times 10^6$) | 0.10525 | $0.0^2 725$ | $0.0^3 923$ |
| Bayesian: McMC ($N = 4 \times 10^6$, Normal) | 0.10744 | $0.0^2 841$ | $0.0^2 118$ |
| (Simulation SD) | $(0.0^3 484)$ | $(0.0^3 186)$ | $(0.0^4 789)$ |
| {Acceptance rate} | {41.9%} | {41.9%} | {41.9%} |
| Bayesian: AMcMC ($N = 4.10^6$, adaptive Student) | 0.10752 | $0.0^2 836$ | $0.0^2 118$ |
| (Simulation SD) | $(0.0^3 332)$ | $(0.0^3 100)$ | $(0.0^4 366)$ |
| {Acceptance rate} | {51.1%} | {51.1%} | {51.1%} |

In terms of these, Welch and Peers (1963) showed in effect that $b$ has a location model $f(b-\beta)$ to second-order accuracy. This has profound implications for possible second order agreement between Bayesian and frequentist methodologies, but Welch and Peers (1963) presented their results in terms of the frequentist approach of obtaining confidence bounds by integrating likelihood with respect to the Jeffreys (1946) prior $i_{\varphi\varphi}^{1/2}(\varphi)d\varphi$. The same result with some greater generality can be obtained by Taylor series expansion of an asymptotic model with scalar variable and parameter; simple reexpressions of variable and parameter show that to second order the model can be written either as a location or as an exponential model; see, for example, Cakmak et al. (1998) and Andrews, Fraser and Wong (2005). A somewhat similar result is available with vector variable and parameter; see Cakmak, Fraser and Reid (1994).

For the case of a location model $f(s-\beta)$ with a flat prior $\pi(\beta) = c$, the expression (20) for $q_B$ simplifies to

$$q_B = \ell_\beta(\beta)\hat{j}_{\beta\beta}^{-1/2}.$$

As $\ell_\beta(\beta; s) = (\partial/\partial\beta)\ell(\beta; s) = -(\partial/\partial s)\ell(s-\beta) = -\ell_{;s}(\beta; s)$ from the location property, and $-\ell_\beta(\beta) = \ell_{;s}(\beta, s) = \varphi$ from the exponential form (27), we obtain through simple algebra that $q_f = q_B$, which implies that the frequentist distribution function is equal to the Bayesian survivor function, as would be expected. In other words, this Welch and Peers (1963) Bayesian–frequentist equality is obtained from the Jeffreys prior for the scalar exponential model, and thus as demonstrated by Cakmak et al. (1994, 1998) has an extension for general asymptotic models. The advantages of this scalar parameter use of the Jeffreys prior has recently been discussed for the discrete binomial distribution context by Brown, Cai and DasGupta (2001).

Now consider the vector exponential model (27) with $p$-dimensional canonical parameter $\varphi$ and $p$-dimensional canonical variable $s$, and suppose that we are interested in a scalar component parameter $\psi(\varphi)$ having reasonable smoothness properties.

The signed likelihood root $r = r_\psi$ in (10) is the primary departure measure and a maximum likelihood departure

$$q_f(\psi) = \text{sign}(\hat{\psi} - \psi) \qquad (31)$$
$$\cdot |\hat{\chi} - \hat{\chi}_\psi|\left\{\frac{|\hat{j}_{\varphi\varphi}|}{|j_{(\lambda\lambda)}(\hat{\theta}_\psi)|}\right\}^{1/2}$$

is the secondary departure measure. For this, $\theta = (\lambda, \psi)$ has been presented as a combination of $\psi$ with a nuisance parameter $\lambda$ which complements the interest parameter $\psi$; in the vector case we should perhaps write the combination in term of row vectors as $\theta' = (\psi', \lambda')$. The scalar parameter $\chi(\theta)$ is a rotated coordinate of $\varphi(\theta)$ that acts as a surrogate for $\psi(\theta)$ and has linearity in terms of $\varphi(\theta)$. Explicit formulas for the surrogate parameter $\chi(\theta)$ and the nuisance information are recorded in the Appendix; some discussion also appears in the next section; the parentheses around $(\lambda\lambda)$ are to indicate that the information has been recalibrated in terms of the new parameterization $\varphi$. All of this is easily accessible numerically, and uses primarily just the typical ingredients of the Bayesian-type approximation.

## 7. THIRD ORDER FOR THE EXAMPLE

We have just described how third-order $p$-values are available to assess a scalar parameter $\psi(\theta)$ in an exponential model. While exponential models are of course quite important, they do represent a very specialized type of model. However, recent likelihood theory has shown that for a general continuous statistical model together with data, there exists a corresponding exponential model that provides highly accurate third-order $p$-values for the original model and data, using the formulas in the preceding section.

For our example in Section 3, the corresponding exponential model with data has the same observed log-likelihood $\ell(\theta) = \log f(y^0; \theta)$ given as (9) and has a nominal reparameterization $\varphi' = (\varphi_1, \varphi_2, \varphi_3)$ given as a row vector; some discussion is given later. See also Davison, Fraser and Reid (2006). The reparameterization is

$$\varphi'(\alpha, \beta, \sigma)$$
$$(32) \qquad = 8\sum_{i=1}^{7}\frac{(\alpha + \beta x_i - y_i^0)/7\sigma^2}{1 + (y_i^0 - \alpha - \beta x_i)^2/7\sigma^2}(1, x_i, d_i^0),$$

and is explained later in detail; here $d^0$ is just the standardized residual vector $(y^0 - \hat{y}^0)/\hat{\sigma}^0$ recorded numerically preceding (35). The corresponding general formulas are recorded at the end of this section.

To obtain the $p$-value for assessing any scalar component parameter, it suffices to treat the observed likelihood as a function of $\varphi$, which of course means explicitly or implicitly that the observed informations needs to be reexpressed or recalibrated in terms of the $\varphi$ parameterization and the maximum likelihood departure needs also to be expressed in the $\varphi$ parameterization.

TABLE 3
*Frequentist p-values for the simple regression example in Section 2, for assessing the values $\beta = 1$, $\beta = 1.5$ and $\beta = 2$: using the third-order formula (30) with (11) and (31); using McMC with normal proposal centered at a value at hand with standard deviation 0.35 (Section 8); using McMC with a centered Student(7) proposal; and using McMC using an adaptive Student proposal (Section 13)*

| Test procedure | p-value | | |
|---|---|---|---|
| | $\beta = 1$ | $\beta = 1.5$ | $\beta = 2$ |
| Frequentist; third order | 0.10525 | $0.0^2 725$ | $0.0^3 923$ |
| Frequentist; McMC ($N = 4.10^6$, Normal) | 0.10832 | $0.0^2 819$ | $0.0^2 118$ |
| (Simulation SD) | $(0.0^3 398)$ | $(0.0^3 109)$ | $(0.0^4 406)$ |
| {Acceptance rate} | {38.0%} | {38.0%} | {38.0%} |
| McMC [$N = 4.10^6$, Student(7)] | 0.10765 | $0.0^2 827$ | $0.0^2 113$ |
| (Simulation SD) | $(0.0^3 196)$ | $(0.0^4 510)$ | $(0.0^4 185)$ |
| {Acceptance rate} | {75.9%} | {75.9%} | {75.9%} |
| AMcMC ($N = 4.10^6$, adaptive Student) | 0.10792 | $0.0^2 823$ | $0.0^2 109$ |
| (Simulation SD) | $(0.0^3 204)$ | $(0.0^4 646)$ | $(0.0^4 264)$ |
| {Acceptance rate} | {81.6%} | {81.6%} | {81.65%} |

The recalibration of the information is obtained from the derivative $\varphi_\theta(\theta) = \partial\varphi/\partial\alpha\,\partial\beta\,\partial\sigma$ of $\varphi$ with respect to the initial parameters as evaluated at the maximum likelihood values. For this with $\psi = \beta$ and $\lambda' = (\alpha, \log\sigma)$, we obtain

$$\varphi_\theta(\hat{\theta}^0) = \begin{pmatrix} 5.7894195 & 1.311060 & -0.3286837 \\ 1.3110600 & 27.288552 & -1.3395559 \\ -0.3286837 & -1.339556 & 12.1900132 \end{pmatrix},$$

$$\varphi_\lambda(\hat{\theta}_{\beta=1}) = \begin{pmatrix} 4.0240689 & -0.3319537 \\ -0.1474505 & -7.5706463 \\ -0.5188019 & 7.5500100 \end{pmatrix}.$$

Using these as scaling matrices along with the matrices (24) gives the recalibrated information determinants:

$$|j_{\varphi\varphi}| = |j_{\theta\theta}||\varphi_\theta|^{-2} = 0.000528345,$$

$$|j_{(\lambda\lambda)}(\hat{\theta}_{\beta=1})| = |j_{\lambda\lambda}(\hat{\theta}_{\beta=1})||\varphi'_\lambda\varphi_\lambda|^{-1} = 0.01844021.$$

The special maximum likelihood departure used in (31) is $\text{sgn}(\hat{\beta} - \beta)|\hat{\chi} - \hat{\chi}_\beta| = -5.602751$. We then use $r_{\beta=1} = -1.574053$ from (11) together with

(33) $$q_{\beta=1} = -0.9483686$$

from (31) to substitute in (30); this gives the third-order $p$-value

(34) $$p_\text{3rd} = 0.10525,$$

which is recorded in Table 3 along with other values including those for testing the parameter values $\beta = 1.5$ and $\beta = 2$.

We can also calculate the third-order frequentist $p$-value $p(\beta)$ for a range of values for $\beta$; for our example, this $p$-value function is plotted in Figure 1 using dots to allow for comparison with the Bayesian $s(\beta)$ obtained in Section 4.

We record now some general thoughts on the reparameterization $\varphi(\theta)$. In the context with independent scalar coordinates, we have

$$\varphi(\theta) = \sum_{i=1}^n \left.\frac{\partial\ell(\theta; y_i)}{\partial y_i}\right|_{y_i^0} \left.\frac{dy_i}{d\theta}\right|_{(y_i^0, \hat{\theta}^0)}.$$

The first factor is a function of $\theta$ that records how the $i$th coordinate influences the likelihood function:

$$\frac{\partial\ell(\theta; y_i)}{\partial y_i} = \frac{\partial}{\partial y_i}\log f_i(y_i; \theta);$$

it is the coordinate gradient of likelihood and can be viewed as a parameter when the observed data values are substituted. For our example we have

$$\left.\frac{\partial\ell(\theta; y_i)}{\partial y_i}\right|_{y_i^0} = 8\frac{(\alpha + \beta x_i - y_i^0)/7\sigma^2}{1 + (y_i^0 - \alpha - \beta x_i)^2/7\sigma^2}$$

and it appears in (32) above. The second factor in (32) is a numerical row vector that records how parameter change near the overall maximum likelihood value affects the $i$th coordinate; it records the sensitivity of the $i$th coordinate to parameter change at the maximum likelihood value. This uses the error $z_i$ as an $i$th coordinate pivot, which with continuity is necessarily one–one equivalent to the distribution function $F_i(y_i; \theta)$; and then with the pivot inverted to express $y_i = y_i(\theta, z_i)$ in terms of $\theta$ and $z_i$ we obtain the derivative

$$\left. \frac{dy_i}{d\theta} \right|_{(y_i^0, \hat{\theta}^0)}$$

for the $i$th coordinate at the data point. For our example with $z_i = (y_i - \alpha - \beta x_i)/\sigma$ we obtain

$$y_i = \alpha + \beta x_i + \sigma z_i^0$$

and then have

$$\frac{\partial y_i}{\partial \alpha} = 1,$$
$$\frac{\partial y_i}{\partial \beta} = x_i,$$
$$\frac{\partial y_i}{\partial \sigma} = z_i.$$

Then for evaluation at $(y^0, \hat{\theta}^0)$ we need the observed standardized residual $\hat{z}^0 = d^0 = d(y^0)$:

$$\begin{aligned} d^0 &= \{d_1(y^0), \ldots, d_7(y^0)\}' \\ &= (0.5614092, -1.1324253, \\ &\quad -0.7667588, 1.2969452, \\ &\quad 0.8640297, -0.2581882, -0.5650118)' \end{aligned}$$

as calculated from

$$(35) \qquad d_i^0 = \frac{y_i^0 - \hat{\alpha}^0 - \hat{\beta}^0 x_i}{\hat{\sigma}^0}.$$

We thus obtain the final vector in $(1, x_i, d_i^0)$ as used in (32).

For some background theory see Fraser and Reid (1993, 1995, 2001) and for an overview of the methodology for the regression context see Fraser, Reid and Wong (2005), Fraser, Wong and Wu (1999).

## 8. THE EXACT $p$-VALUES AND $s$-VALUES

Higher-order $p$-values and higher-order posterior survivor $s$-values are usually validated by simulations, by verifying that the large sample distribution in each case is close to the uniform$(0, 1)$ distribution. The formulas, however, have been developed in the context of a conditional model: in the Bayesian context it is conditional on the full data, and in the frequentist context it is conditional on data indicators of statistical precision, which typically are given by exact or approximate ancillaries that reflect structure and continuity of the model with respect to the variable and the parameter.

For a general frequentist context, the approximate ancillaries for third-order inference are well defined theoretically but are only needed near the observed data point and typically are only available near the observed data. In such contexts this can make a full conditional validation unattainable as typically there is no accessible information concerning other conditioned points, beyond the tangent direction at the data. Our example, however, has special linear and transformation properties that do provide an exact conditioning variable, here $d(y)$, and thus an exact conditional distribution; some background and details are recorded in the Appendix at point (i). Again as in Section 2 we describe the model in terms of the convenient least squares coordinates $(a, b, s)$, and then record the density for just the standardized or null case with $\alpha = 0$, $\beta = 0$, $\sigma = 1$; the conditional density for $(a, b, s)$ is

$$(36) \qquad g(a, b, s) = c \prod_{i=1}^{7} h\{a + bx_i + sd_i\} s^4,$$

where as before $h(z)$ is the Student(7) density; this is derived in Fraser (1979) and discussed briefly in Fraser (2004). Note that we could also have used maximum likelihood variables, as the error distribution is free of the parameter, but the least squares variables have convenient simplicity; the nonnull conditional density is then available directly as

$$\sigma^{-3} g\{\sigma^{-1}(a - \alpha), \sigma^{-1}(b - \beta), \sigma^{-1}s\}.$$

The null and nonnull distributions can also be expressed directly in terms of the observed likelihood function $L^0(\alpha, \beta, \sigma)$, by simple change of argument; for details, see the Appendix at point (i).

More generally, for a regression model $y = X\beta + \sigma z$ where $z$ has error density $f(z) = \prod_{i=1}^{n} g(z_i)$ and $X$ is $n \times r$ with full column rank, the conditional null density for the least squares $(b, s)$ given the observed value of the residual vector $d = s^{-1}(y - Xb)$ is

$$(37) \qquad cf(Xb + sd^0) s^{n-r-1},$$

which is the original density reexpressed in terms of the new variables coupled with a Jacobian scaling factor with power equal to the effective number of coordinates that are conditioned. For sample simulations, next to be discussed, we will, however, switch from $s$ to $\log s$ to obtain an unbounded range, with corresponding effect on the density expressions (36) and (37).

To assess the accuracy of the third-order Bayesian values from Section 5 or the third order frequentist $p$-values from Section 6, we will use large-scale computer simulations from the posterior density (15) or from the precision-based conditioned density (36) or (37). While for our example with three coordinates, numerical integration would be feasible, we choose the more flexible and generally available Markov chain Monte Carlo (McMC) sampling procedure.

For the McMC sampling we will refer to the distribution to be sampled as the target distribution or target density and use the notation $g(y)$; in our case and frequently in general such target distributions come to us unnormed, that is, we do not know the value of the integral $\int g(y)\,dy$. We describe a procedure for successively obtaining sample values $y_1, y_2, \ldots$. In particular, with a given sample value $y_i$ in hand, we sample from a normal distribution located at that sample value with coordinate standard deviation, here say 0.35, to obtain a possible next sample value; this normal distribution is called a Gaussian proposal. We then use a ratio of likelihood at the possible new value to likelihood at the value in hand to decide whether the next value $y_{i+1}$ is to be the just-obtained trial sample value or is to be a repeat of the value in hand; the likelihood ratio is called the Metropolis–Hastings criterion. We are thus using a random walk Metropolis (RWM) algorithm with a Gaussian proposal distribution to generate a sequence $(y^1, \ldots, y^N)$ of points where $y$ here refers to the variable in the target distribution being sampled, that is, the posterior (15) or the modified version of the three-dimensional distribution (36) or (37). The limiting distribution of the sequence approximates the distribution of the target but does have serial correlation that complicates the estimation of the effective simulation sampling variance. An alternative and convenient proposal distribution is the uniform proposal, a uniform distribution centered again at the value in hand with range here, say, 1.50. For a recent overview see Robert and Casella (2004).

To estimate the true $p$-value based on the observed $t$-departure from our original data set, we then check for each sample point whether $t$ in (2) calculated from a simulated $y$ is less than the observed $-1.445634$; the simulated exact $p$-value is obtained as the proportion satisfying the inequality.

We will also report the estimated simulation standard deviation; results are recorded in Table 3 together with the third-order $p$-value from Section 7 and some other values. The simulation size was $N = 4{,}000{,}000$; in the sample sequence we would dump 50 values, then retain 950 values and repeat this pattern. From this sampling pattern we were able to obtain an estimate of the simulation standard deviation, using the 4000 repeats of sample means from batches of 950; for some details, see the Appendix at point (ii). The table also records $p$-values for testing $\beta = 1.5$ and $\beta = 2.0$ using corresponding observed values $t^0 = -3.669685$ and $t^0 = -5.893737$.

## 9. PRECISION INFORMATION AND BAYESIAN-FREQUENTIST AGREEMENT

With continuous parameters and theory based nominally on increasing amounts of data we have noted that $p$-values for scalar parameters are available with third-order accuracy. Similarly, upper tail posterior values or $s$-values for scalar parameters are also available with third-order accuracy assuming of course the acceptability of the prior. In the default prior community, it seems acknowledged that the choice of sensible prior needs to be based on the parameter of interest, in other words, targeted on the parameter of interest; the development of targeted default priors will be examined separately.

For the frequentist approach we have noted that third-order methods relate implicitly to conditioning on precision information obtained from the data; and the conditioning effectively reduces the dimension of the active variable to the dimension of the parameter; for some recent discussion see Casella, DiCiccio and Wells (1995) and Fraser (2004).

In a paper presented at the International Workshop on Objective Bayesian Methodology at the University of Valencia on June 13, 1999, one of the present authors discussed strong matching, defined to be the effective equivalence of the Bayesian $s$-value and the frequentist $p$-value. The issue in the Bayesian context of having the choice of prior also reflect conditioning on precision information provided by the model and data was mentioned by the presenter and independently by the discussant T. Severini of Northwestern University. The issue centered on a model with scalar parameter $\theta$ and a data precision indicator $a$ such that the actual measurement of $\theta$ was made by the submodel $f_1(y; \theta)$ if $a = 1$ and by submodel $f_2(y; \theta)$ if $a = 2$; the data indicator had a fixed distribution equivalent to the toss of a fair coin. This random choice of measurement model was proposed by Cox (1958).

Suppose that the model $f_1(y;\theta)$ has information $i_1(\theta)$ and the model $f_2(y;\theta)$ has information $i_2(\theta)$: the information for the composite model is then $i(\theta) = \{i_1(\theta) + i_2(\theta)\}/2$. The Jeffreys prior gives the posterior

$$(38) \qquad p(\theta|y,a) = cf(y,a;\theta)i^{1/2}(\theta),$$

and it has a second-order location relationship with a reexpressed $\theta$ (Welch and Peers, 1963). This Bayesian posterior distribution is of course conditional on the observed data, but the indicated choice of prior does not reflect the information that the data has identified the model type that actually made the measurement. If the corresponding information is used to assist the determination of the prior, then the posterior, with reference to Jeffreys, would be

$$(39) \qquad p(\theta|y,a) = cf(y,a;\theta)i_a^{1/2}(\theta),$$

where $a$ has its observed value; see also Fraser (2004).

Clearly (38) and (39) differ whenever $i_1(\theta)$ differs from $i_2(\theta)$. Should the default Bayesian allow the default or invariant prior to depend on information provided by the data?

At the workshop there was some acknowledgment of a place for such precision information in the choice of default prior. If such conditioning is accepted among Bayesians and frequentists, then agreement to third order is possible: the $p$-values are equal to the $s$-values and the professional disagreement would seem to vanish. What then seems clear in the context of continuous parameters and modest regularity, is that the frequentist $p$-values and the Bayesian $s$-values become equal if the frequentist accepts conditioning on observed precision information and the Bayesian suitably targets his default prior, responding to similar information.

## 10. BAYESIAN POSTERIOR MEANS AND VARIANCES

We have been discussing the use of model precision information for the choice of a default prior. Now suppose we take a prior for convenience or expediency or otherwise, and wish to obtain some general posterior characteristics such as posterior means and variances. Means and standard deviations can often provide a convenient summary for purposes of inference, both frequentist and Bayesian. The data-accretion techniques apply to likelihood of course, and consequently to weighted likelihood as given by a posterior. Accordingly we discuss briefly how these large-sample-type techniques can help in the Bayesian context.

Consider a component scalar parameter $\psi(\theta)$ and suppose we want just its mean and variance

$$M = E\psi(\theta) = \int c\psi(\theta)L(\theta)\pi(\theta)\,d\theta,$$

$$V = E\{\psi(\theta) - M\}^2$$
$$= \int c\{\psi(\theta) - M\}^2 L(\theta)\pi(\theta)\,d\theta,$$

where $c$ here is the norming constant for the posterior distribution. We have of course the option of extensive McMC simulations. We first, however, examine higher-order likelihood-based methods that can be applied or adapted to this purpose.

From Section 2 and assuming we have a convenient nuisance parameterization $\lambda$, we obtain

$$(40) \qquad \bar{f}(\psi) = ce^{-r_\psi^2/2}\left\{\frac{|\hat{\hat{j}}_{\theta\theta}|}{|j_{\lambda\lambda}(\hat{\theta}_\psi)|}\right\}^{1/2}\frac{\pi(\hat{\theta}_\psi)}{\pi(\hat{\theta})}$$

as the third-order posterior density approximation when renormalized, and

$$(41) \qquad \bar{F}(\psi) = 1 - \Phi\left(r_\psi - r_\psi^{-1}\log\frac{r_\psi}{q_\psi}\right)$$

as the third-order posterior distribution function approximation; the signed likelihood root $r_\psi$ is given by (10) in Section 3 and the adjusted maximum likelihood departure $q_\psi$ by (20) in Section 4. With third-order accuracy for $\bar{f}$ and $\bar{F}$ we have third-order accuracy available in principle for obtaining the means and variance. A generating-type function to accomplish this would be appealing but seems inaccessible. Using the distribution function directly, however, we do have the following reexpressions:

$$(42) \qquad E(\psi) = \int_0^\infty \{1 - F(\psi) - F(-\psi)\}\,d\psi,$$

$$(43) \qquad E(\psi^2) = \int_0^\infty \{1 - F(\psi) + F(-\psi)\}2\psi\,d\psi.$$

As part of the usual computation of quantities such as the distribution function $\bar{F}(\psi)$, a familiar numerical practice is to evaluate the quantity at equally spaced points, say

$$\ldots, \psi_0 - 2\delta, \psi_0 - \delta, \psi_0, \psi_0 + \delta, \psi_0 + 2\delta, \ldots$$

taken about some convenient central value $\psi_0$ using a small value for $\delta$. Let

$$\ldots, F_{-2}, F_{-1}, F_0, F_1, F_2, \ldots$$

designate such distribution function values, which can conveniently be stored in a file. We then have that

$$E(\psi - \psi_0) = \delta\{\cdots + F_{-2} + F_{-1} \\ + (1 - F_1) + 2(1 - F_2) + \cdots\},$$

$$E(\psi - \psi_0)^2 = 2\delta^2\{\cdots + 2F_{-2} + 1 \cdot F_{-1} \\ + 1 \cdot (1 - F_1) + 2(1 - F_2) + \cdots\}$$

are available immediately by cumulative sums through the stored file; we then directly obtain $E(\psi)$ and $V(\psi)$.

For our regression example as discussed in Sections 2 and 3, we have used a conventional default prior for the Bayesian considerations in Section 4. The corresponding posterior survivor function for $\beta$ was recorded as Figure 1 in Section 3. We processed the related file by the above summation formulas allowing for the fact that $\bar{F}(\beta) = 1 - s_B(\beta)$ and obtained the values in Table 4, column I.

We also differenced the distribution function values to get density values and used them for ordinary numerical integration to obtain the mean and variance; the results are recorded in column II.
As a more direct numerical approach, we used the estimated density $\bar{f}$ as given by (40) to obtain an alternate value by ordinary numerical integration; the results are recorded in column III.

Finally we used the Markov chain Monte Carlo methods as described in Section 7 to simulate these posterior means and variances; the simulation size was $N = 4{,}000{,}000$. The results for the normal sampling proposal are recorded in column IV of Table 4; the results for the uniform sampling proposal were very close. We note the high accuracy of the third-order procedures relative to the simulated exact, but do not here attempt a detailed comparison of I, II, III.

## 11. SOME THOUGHTS ON McMC

Consider a target density $g(y)$ that we wish to sample from: for the example the target $g(y)$ is given by (14) for the Bayesian approach and by (36) for the conditional frequentist case. In both cases the $g(y)$ is unnormed; it is a relative density function. The sampling difficulty for simulations is typically due to the fact that the target density is not a product of independent variables with the related ease of sampling coordinate by coordinate. For notation we will assume that $g(y)$ is normed, but this will not be used other than to facilitate the discussion.

In this section we briefly discuss the McMC methodology from a statistical rather than probabilistic point of view and accordingly use notation that is more statistical. This seems particularly appropriate in the present context of comparing statistical inference from the Bayesian and frequentist approaches: both give unnormed density functions that are conditional. The large sample techniques give highly accurate third-order results; these can be assessed by McMC and improvement can be obtained by increasing the simulation size $N$.

The theme behind the Markov chain Monte Carlo procedure is to use an accessible density function, say $f(x|y)$, to produce a possible value $x$ for the next value in a sample sequence, based of course on the most recent value, say $y$. This accessible density is typically taken to have an amenable product form with independent coordinates. In the McMC sampling process with values $y^1, \ldots, y^n$ in hand, we sample from the proposal density $f(x|y^n)$ to obtain a candidate $x$ for the next sample value. This candidate will either be accepted with an acceptance probability $A(x|y^n)$ with the result that $y^{n+1}$ is set equal to $x$, or be rejected with complementary probability $1 - A(x|y^n)$ with the result that $y^{n+1}$ is set equal to $y^n$ which is a repeat of the value in hand. The acceptance probability is often taken to be a Metropolis–Hastings ratio (Metropolis, 1953; Hastings, 1970), now to be described.

For discussion we let $g(x)$ and $f(x|y)$ designate probabilities of being in a neighborhood of a point $x$. Of course we should properly write $g(x)\triangle$ and $f(x|y)\triangle$, where $\triangle$ is a small volume element at the point $x$, but all the $\triangle$'s will cancel and expressions are easier without them.

TABLE 4
*Third-order posterior mean, variance and standard deviation by method I, II, III; validation by McMC with $N = 4 \cdot 10^6$*

| Computation method | I | II | III | McMC |
|---|---|---|---|---|
| Mean = $E(\beta)$ | 0.67642 | 0.67639 | 0.67208 | 0.67172 |
| Simulation SD | | | | $(0.0^3 550)$ |
| Variance = $V(\beta)$ | 0.08096 | 0.08101 | 0.08615 | 0.08436 |
| Simulation SD | | | | $(0.0^3 665)$ |
| SD = $SD(\beta)$ | 0.28453 | 0.28463 | 0.29352 | 0.28642 |
| Simulation SD | | | | $(0.0^3 763)$ |

Consider two sample space points $a$ and $b$, and how a next sample point might be a transition or a repeat among these points. If we are at a first time point with data value $a$ and sample from the proposal $f(x|a)$ when the target to sample from is $g(x)$, then the likelihood ratio

$$L(b) = \frac{g(b)}{f(b|a)}$$

records how things ideally should be scaled to agree with the target at $b$. Alternatively, if we are at the same first time point with data value $b$ and sample from the proposal $f(x|b)$ when the target to sample from is $g(x)$, then the likelihood ratio

$$L(a) = \frac{g(a)}{f(a|b)}$$

records how things ideally should be scaled to agree with the target at $a$. The ratio of these likelihoods is called the Metropolis–Hastings ratio,

$$MH(b|a) = \frac{L(b)}{L(a)}$$
$$= \frac{g(b)/f(b|a)}{g(a)/f(a|b)} = \frac{g(b)f(a|b)}{g(a)f(b|a)};$$

it gives us the mechanism to adjust the transition from $a$ to $b$ to give conformity to the target density $g(y)$; its reciprocal addresses the transition from $b$ to $a$. Accordingly, the acceptance probability for going from $a$ to $b$ is taken to be $MH(b|a)$ but capped at the maximum 1 possible for a probability:

$$A(b|a) = \overline{MH}(b|a),$$

where the bar indicates the capping: $\bar{A} = \min(A, 1)$. The capping can of course give a shortfall in transitions from $a$ to $b$ but we see that the related rejection and repeat of the preceding value precisely compensates.

Consider the typical case where the proposal $f(x|y)$ does not duplicate the target $g(x)$. And without loss of generality, consider a pair of points $a, b$ where $L(a) \geq L(b)$ or

$$MH(b|a) = A(b|a) = \frac{g(b)f(a|b)}{g(a)f(b|a)} \leq 1.$$

Suppose the probabilities for the sampling process are correct at some possible point in time, that is, they are equal to $g(a)$ and $g(b)$ corresponding to $a$ and $b$ for that time point. Then suppose we consider transitions within the pair $\{a, b\}$. Concerning a transition going from $a$ to $b$, the probability of being at $a$ and going to $b$ is

$$g(a)f(b|a)A(b|a) = g(b)f(a|b);$$

this represents a probability increase at $b$ and probability loss at $a$. Concerning a transition from $b$ to $a$, while noting that $A(a|b) = 1$, the probability of being at $b$ and going to $a$ is

$$g(b)f(a|b)A(a|b) = g(b)f(a|b);$$

this represents a probability increase at $a$ and probability loss at $b$. We note that the two probability movements cancel each other and thus the probabilities at $a$ and $b$ are maintained; we do note, however, that the rejection probability $1 - A(b|a)$ represents a loss of new sampling information. One can thus view the acceptance probability as an effective adjustment of the proposal $f(x|y)$ to yield proper transitions between pairs of points so as to accomplish what is prescribed by the target $g(y)$.

## 12. ASYMPTOTIC McMC

For our example, the large sample likelihood-based methods gave us a Bayesian analysis in Section 4 and a frequentist analysis in Section 6. In both cases, we used Markov chain Monte Carlo methods for validation: in the first case we sampled the unnormed posterior $\pi(\theta)L(\theta)$ given by (14) which is a conditional distribution given the data; in the second case we sampled an unnormed sample space conditional distribution given by (36) which is conditional on observed precision information. Of course the example is sufficiently low dimensional that numerical integration could have been used, but McMC is easier to implement and readily extends to larger and more complicated sample spaces and target densities $g(y)$. We now examine how we can make use of the asymptotic form of the target distribution to give a more efficient version of the proposal distribution. For our example, the Metropolis–Hastings acceptance rates were approximately 38% for the normal proposal and 25% for the uniform proposal, and both yielded reasonable convergence rates. We now investigate ways to smartly increase this acceptance rate, by introducing a proposal density that generates wiser moves and thus improves the precision of the McMC sampling process.

For our asymptotic context we have that the unnormed density $g(y)$ has a maximum density value at a point which we designate as $\hat{y}$, and has a negative Hessian designated as $\hat{j} = -\partial^2 \log g(y)/\partial y\, \partial y'$ as evaluated at the maximum $\hat{y}$. We now assume that the variable $y$ has dimension, say, $d$, and investigate the choice of an expedient proposal $f(x|y)$. In Section 7, we used a proposal that was a product of independent normals and another that was a product of independent uniforms:

$$(44) \qquad f(x^{n+1}|y^n) = \prod_{i=1}^{d} h(x_i^{n+1}|y_i^n),$$

where $h$ is either normal or uniform centered at $y^n$, with scaling chosen pragmatically; in the present case $x$ and $y$ have dimension $d$ with coordinates $x_1, \ldots, x_d$ and $y_1, \ldots, y_d$. Of course a proposal that mimics the target $g(y)$ would have advantages in efficiency, giving candidate sample values that tend to agree with the target $g(y)$ and thus have less loss.

A simple choice for the proposal $f(x|y)$ is the $N(\hat{y}; \hat{j}^{-1})$ distribution which does not depend on the preceding value. Such a multivariate normal is available in many computing packages for random sampling and has here both the same point of maximum value and the same local scaling as the target distribution. The normal, however, has short tails and thus in sampling would often neglect the tails of a target distribution $g(y)$, with loss of efficiency perhaps serious in some contexts.

A more refined choice is the multivariate Student distribution with degrees of freedom chosen pragmatically to give longer tails (see Brazzale, 2000); this proposal is purely for the McMC simulations and does not relate to the objective of interest, the $p$-value, although it does affect the McMC simulations, as we will see. A canonical version of this Student distribution with degrees of freedom, say, $f$ is designated Student$_f(0; I)$ and has density

$$(45) \qquad h(T) = \frac{\Gamma((f+d)/2)}{\pi^{d/2}\Gamma(f/2)} \cdot (1 + T_1^2 + \cdots + T_d^2)^{-(f+d)/2};$$

sample values for this can be obtained as

$$T' = \left(\frac{z_1}{\chi_f}, \ldots, \frac{z_d}{\chi_f}\right),$$

where the $z_i$ are independent standard normal and $\chi_f$ is a chi variable with $f$ degrees of freedom, both easily accessible in computer packages. The negative Hessian of the canonical log-density from (45) is $(f+d)I$ and for use in the present context would need to be adjusted by scaling and also of course by location to give the desired location and Hessian to match the target. Accordingly, we take the modified proposal $f(x|y)$ replacing (44) to be the Student$_f\{\hat{y}, (f+d)\hat{j}^{-1}\}$ with values available as

$$\begin{aligned}
x &= \hat{y} + (f+d)^{1/2}\hat{j}^{-1/2}\,T \\
(46) \qquad &= \hat{y} + (f+d)^{1/2}w/\chi_f \\
&= \hat{y} + W/\chi_f,
\end{aligned}$$

where $T$ designates a vector from the canonical Student$_f(0, I)$, $w$ designates a value from the multivariate normal $MV(0, \hat{j}^{-1})$, $W$ designates a value from the $MN(0, (f+d)\hat{j}^{-1})$ and $\hat{j}^{1/2}$ is a suitable square root matrix of $\hat{j}$. A pragmatic choice for the degrees of freedom $f$ could allow for thicker tails and provide improved sampling coverage of extremes. For our example we simplistically chose the degrees of freedom $f$ to be the degrees of freedom 7 that was used originally to generate the individual coordinates.

We then applied the McMC procedure using the Metropolis–Hastings ratio and sampled $N = 4{,}000{,}000$ times using the dump 50, keep 950 procedure as described earlier; we then calculated the proportion of values with

$$t \leq -1.445634,$$

or equivalently with

$$b/s \leq -0.122178.$$

We obtained the $p$-value $p = 0.10765$ with simulation $SD = 0.000196$, along with an acceptance rate of 76%; see Table 3. Values are also recorded for testing $\beta = 1.5$ and $\beta = 2$. Our view verified so far is that the more the proposal mimics the target, the higher the acceptance rate will be. To obtain high accuracy, very large values of $N$ are needed so any increase in efficiency has merit. We discuss this briefly in the final discussion section.

## 13. ADAPTIVE McMC

The use of a RWM sampling proposal $f(x|y)$ as in Section 7 is in its nature adaptive, as it samples near the most recent sample value. We modify this adaptive procedure by having the proposal mimic the target $g(\cdot)$, that is, by having the same shape at the maximum and the same drop-off to the current value $y$ in hand. We do this by centering and shaping the proposal as in the preceding section, but also by determining the degrees of freedom $f$ to duplicate the drop-off

$$(47) \quad \frac{g(y)}{g(\hat{y})} = \left\{1 + \frac{(y-\hat{y})'\hat{j}(y-\hat{y})}{f+d}\right\}^{-(f+d)/2}$$

from the maximum to the current point $y$; we then have that the Student proposal has the same maximizing value and has the same Hessian as the target, but now also has the degrees of freedom to duplicate the tail thickness at the current point in hand. For pragmatic reasons we take $f$ to be the nearest integer to the solution of (47) but restrict it to the range from, say, the Cauchy with $f=1$ to the near normal with $f=50$. If we let $r^2 = 2\log\{g(\hat{y})/g(y)\}$ be the target likelihood ratio quantity and $Q^2 = (y-\hat{y})'\hat{j}(y-\hat{y})$ be the quadratic departure for the Student, we can solve for $f + d = \overline{f}$ using

$$(48) \quad \overline{f} \log\left(1 + \frac{Q^2}{\overline{f}}\right) = r^2$$

by a simple scan of integer values for $f$ in $\{1, 2, \ldots, 50\}$.

This adaptive McMC then proceeds as follows: if the $i$th sample value $y^i = y$, we solve for an integer $\overline{f}$ and then $f$ using (48) and (47) and obtain a trial value $x$ for the next observation by sampling from $f(x|y)$ taken to be the Student$_f(\hat{y}, (f+d)\hat{j}^{-1})$ distribution using one of the data generation methods in (46).

For the example, we now apply this adaptive procedure to the conditional distribution (36) in Section 7 and examine the proportion of values with

$$t \leq -1.445634$$

or equivalently with

$$b/s \leq -0.122178.$$

With $N = 4{,}000{,}000$ and using the dump 50, keep 950 procedure as before we obtain $p = 0.10792$ with $SD = 0.000204$, along with an acceptance rate of 82%, substantially more than with preceding methods. This is recorded in Table 3 together with values for assessing $\beta = 1.5$ and $\beta = 2$.

## 14. CONTROVERSIAL EXAMPLE: BEHRENS–FISHER

**(with Ye Sun, York University)**

In a recent study of controversial examples in statistics (Fraser, Wong and Sun, 2007), extensive simulations were performed on some recent procedures for the Behrens (1929)–Fisher (1935) statistical problem. This problem concerns a sample of $n_1$ from a Normal($\mu_1, \sigma_1$) and a sample of $n_2$ from a Normal($\mu_2, \sigma_2$) and addresses inference for the difference $\delta = \mu_1 - \mu_2$ of the population means. The statistical model is simple, just two normals, but clearcut procedures for inference have been elusive.

Fisher (1935), following Behrens (1929), suggested that the confidence distribution for $\mu_1$ be convolved with the confidence distribution for $\mu_2$ to target the difference $\delta = \mu_1 - \mu_2$. This combining of confidence distributions ran contrary to statistical practice at the time and evoked an extensive literature response which we do not explore here.

Jeffreys (1961) recommended the use of the prior

$$\sigma_1^{-1}\sigma_2^{-1}\,d\mu_1\,d\sigma_1\,d\mu_2\,d\sigma_2,$$

which is the combination of the right invariant priors for the two normal models. Such right invariant priors are common priors for default Bayesian analysis; also the right invariant prior for a normal model can be seen to reproduce Fisher's confidence distribution for the corresponding mean.

Ghosh and Kim (2001) proposed a second-order default prior

$$\sigma_1^{-3}\sigma_2^{-3}(\sigma_1^2/m + \sigma_2^2/n)\,d\mu_1\,d\mu_2\,d\sigma_1\,d\sigma_2,$$

which has somewhat the form of a weighted average of the two component right invariant priors.

The signed likelihood ratio (10) examined in Section 3 can provide first-order $p$-values and confidence intervals. The $\beta$-level confidence interval for the difference $\delta$ in means has the form

$$(\delta : z_{-\alpha/2} < r_\delta < z_{\alpha/2}),$$

where $(z_{-\alpha/2}, z_{\alpha/2})$ is a $\beta = 1 - \alpha$ interval for the standard normal and $r_\delta$ is the signed likelihood ratio (10) for assessing $\delta$.

The third-order likelihood methods in Section 6 use the signed likelihood ratio $r_\delta$ together with the maximum likelihood departure $q_\delta$ formula (31), and then combine them using Barndorff-Nielsen's (1991) formula (30) to obtain an $r_\delta^*$ for assessing the Behrens–Fisher $\delta$. The corresponding $\beta$-level confidence interval is

$$(\delta : z_{-\alpha/2} < r_\delta^* < z_{\alpha/2}).$$

These methods were compared in Fraser, Wong and Sun (2007) using a simulation size of $N = 10{,}000$. The third-order methods generally performed well, especially with increasing sample size.

TABLE 5
*For a simulation size of $N = 10,000,000$, the table records the percentage of simulation cases where the true value was left or right of the confidence interval*

| Method | 99% CI | | 95% CI | | 90% CI | |
|---|---|---|---|---|---|---|
| | Outside left | Outside right | Outside left | Outside right | Outside left | Outside right |
| Target value | 0.50% | 0.50% | 2.50% | 2.50% | 5.00% | 5.00% |
| Jeffreys | 0.009% | 0.010% | 0.245% | 0.245% | 0.958% | 0.960% |
| Ghosh and Kim | 0.022% | 0.023% | 0.543% | 0.545% | 2.027% | 2.028% |
| Likelihood ratio | 3.884% | 4.421% | 9.718% | 9.247% | 13.597% | 14.142% |
| Third order | 0.402% | 0.401% | 2.021% | 2.023% | 4.045% | 4.043% |
| (2 $\widehat{SD}$ limits) | ($\pm 0.002\%$) | ($\pm 0.002\%$) | ($\pm 0.005\%$) | ($\pm 0.005\%$) | ($\pm 0.007\%$) | ($\pm 0.007\%$) |

The target value is the corresponding confidence value; the methods are the Bayesian Jeffreys and Ghosh and Kim and the frequentist likelihood ratio and the third order.

For presentation here we chose the smallest possible sample sizes $n_1 = n_2 = 2$ and the equal variance case and increased the simulation size to $N = 10,000,000$. Then for central confidence intervals at levels 99%, 95%, 90% we calculated the percentage of cases with true parameter value on the left side and on the right side of the confidence interval; the results are recorded in Table 5, where we also record the estimated simulation limits.

The results address a most extreme case of the Behrens–Fisher problem: samples of size $n_1 = n_2 = 2$. The third-order performance seems reasonably close to the target. It does, however, deviate by more than the simulation limits would suggest; but it does represent a substantial improvement over available procedures.

## 15. DISCUSSION

We have surveyed inference procedures for obtaining frequentist $p$-values and Bayesian posterior survivor $s$-values, as well as the corresponding confidence intervals and posterior intervals. Our emphasis has been on the use of higher-order likelihood methods to obtain increased accuracy and we have verified the increased accuracy with extensive McMC simulations.

To motivate the presentation of the procedures we have used a very simple linear model but with nonnormal errors. The example does have an appropriate default prior so the frequentist and Bayesian methods are comparable.

For a more complex example we have reported on extensive simulations for the most extreme case of the Behrens–Fisher problem, an example that is simple in the sense of involving only normal samples but complex in its long-standing history of defying both frequentist and Bayesian theoretical approaches. The higher-order methods lead to $p$-values that quite accurately assess the difference in means, the typical parameter of interest, and from simulations outperform available Bayesian methods.

We have also examined McMC methods from a statistical viewpoint and illustrated them by extensive assessments of higher-order likelihood methods.

In brief we have found that higher-order likelihood using MLEs and observed information can yield the precision of 4 million simulation steps given a suitable statistic. In addition they provide focused accuracy by precisely separating information on almost any scalar parameter chosen as of interest. Various examples illustrating the theory are also included with the references.

## APPENDIX

**(i) Regression conditional distribution.** For the regression model $y = X\beta + \sigma z$ with error density $f(z) = \prod_{i=1}^{n} g(z_i)$, we can examine how parameter change affects the $n$ coordinates $y_i$ and how continuity determines a conditional distribution having dimension equal to that of the parameter. Convenient coordinates $(b, s)$ corresponding to $(\beta, \sigma)$ are available from least squares or maximum likelihood (see, e.g., Fraser, 1979, 2004); in either case we have

$$b(y) = \beta + \sigma b(z),$$
$$s(y) = \sigma s(z),$$

and then have the standardized residual vector

$$d(y) = s^{-1}(y)\{y - Xb(y)\}$$
$$= s^{-1}(z)\{z - Xb(z)\} = d(z).$$

It follows with observed data $y^0$ that $d(z) = d(y^0)$ which then implies that the appropriate model should be conditional. Routine calculations (Fraser, 1979) then give the null distribution

$$g(b,s)db\,ds = c\prod_{i=1}^{n} g(X_i b + s d_i^0) s^{n-r-1}\,db\,ds,$$

where $X_i$ is the $i$th row of $X$ and $d_i^0$ is the $i$th element of the observed standardized residual $d(z) = d(y^0)$; the nonnull distribution for $\{b(y), s(y)\}$ is then

$$g(b,s;\beta,\sigma)\,db\,ds$$
$$= c\prod_{i=1}^{n} g[\sigma^{-1}\{X_i(b-\beta) + s d_i^0\}]\left(\frac{s^{n-r-1}}{\sigma^n}\right)db\,ds.$$

This can be rewritten directly in terms of the observed likelihood $L^0(\beta,\sigma;y^0) = cf(y^0;\beta,\sigma)$ as

$$g(b,s;\beta,\sigma)\,db\,ds = L^0(\beta_*,\sigma_*)\frac{db\,ds}{s^{r+1}},$$

where

$$\beta_* = b^0 + \frac{s^0}{s}(\beta - b),$$
$$\sigma_* = \frac{s^0}{s}\sigma.$$

**(ii) Simulation standard deviation.** In a Bernoulli sequence the observed proposition $\hat{p}$ has a standard deviation $(pq/N)^{1/2}$, which is bounded by $1/2N^{1/2}$. An McMC sequence will typically have serial correlations; accordingly we worked in batches of $B = 1000$ and dropped the first 50 and retained the remaining 950 in each batch. We tested and found the sequence of $N_B = 4{,}000{,}000/B = 4000$ batch means to be essentially free of correlation. We calculated the usual standard deviation $s$ of the batch means and then obtained an upper bound estimate $s/N_B^{1/2}$ for the standard deviation of the overall mean. This can be fine-tuned for probabilities away from $1/2$ by using $\hat{p}$ in place of $1/2$ in the usual binomial variance formula.

**(iii) The surrogate for $\psi(\theta)$.** The rotated $\varphi$ coordinate is obtained using a coefficient vector $a$ applied to the $\varphi$-vector,

$$(49) \qquad \chi(\theta) = a'\varphi(\theta) = \frac{\psi_{\varphi'}(\hat\theta_\psi)}{|\psi_{\varphi'}(\hat\theta_\psi)|}\varphi(\theta);$$

the row vector $a'$ multiplying $\varphi(\theta)$ is the unit vector version of the gradient $\psi_{\varphi'}(\hat\theta_\psi)$ and is obtained by evaluating

$$\psi_{\varphi'}(\theta) = \frac{\partial\psi(\theta)}{\partial\varphi'}$$
$$= \frac{\partial\psi(\theta)}{\partial\theta'}\left(\frac{\partial\varphi(\theta)}{\partial\theta'}\right)^{-1}$$
$$= \psi_{\varphi'}(\theta)\varphi_{\theta'}^{-1}(\theta)$$

at $\hat\theta_\psi$, and then normalizing; this gives a unit vector perpendicular in the $\varphi$ coordinates to $\psi\{\theta(\varphi)\}$ at $\hat\varphi_\psi$. The use of the unit vector in (49) produces a rotated coordinate of $\varphi(\theta)$ that agrees with $\psi(\theta)$ at $\hat\theta_\psi$ in the sense of being first derivative equivalent to $\psi(\theta)$ at the point $\hat\theta_\psi$.

**(iv) Information determinants.** The information determinants are recalibrated to the $\varphi$ parameterization

$$|\hat j_{\varphi\varphi}| = |\hat j_{\theta\theta}||\varphi_\theta(\hat\theta)|^{-2}$$
$$(50) \qquad |j_{(\lambda\lambda)}(\hat\theta_\psi)| = |j_{\lambda\lambda}(\hat\theta_\psi)||\varphi_{\lambda'}(\hat\theta_\psi)|^{-2}$$
$$= |j_{\lambda\lambda}(\hat\theta_\psi)||X|^{-2},$$

where the right-hand $p \times (p-1)$ determinant $|X| = |X'X|^{1/2}$ uses $X = \varphi_{\lambda'}(\hat\theta_\psi)$ and in the regression context records the volume on the regression surface as a proportion of volume for the regression coefficients.